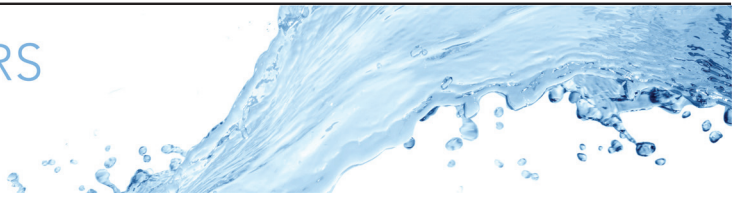

# Stochastic modelling of the instantaneous velocity profile in rough-wall turbulent boundary layers

**Roozbeh Ehsani**[1,2], **Michael Heisel**[3], **Jiaqi Li**[1,4], **Vaughan Voller**[1,2], **Jiarong Hong**[1,4] **and Michele Guala**[1,2,†]

[1]Saint Anthony Falls Laboratory, University of Minnesota, Minneapolis, MN 55414, USA

[2]Department of Civil, Environmental, and Geo- Engineering, University of Minnesota, Minneapolis, MN 55455, USA

[3]School of Civil Engineering, University of Sydney, Darlington 2008, NSW, Australia

[4]Department of Mechanical Engineering, University of Minnesota, Minneapolis, MN 55455, USA

---

The statistical properties of uniform momentum zones (UMZs) are extracted from laboratory and field measurements in rough wall turbulent boundary layers to formulate a set of stochastic models for the simulation of instantaneous velocity profiles. A spatiotemporally resolved velocity dataset, covering a field of view of $8 \times 9\,\mathrm{m}^2$, was obtained in the atmospheric surface layer using super-large-scale particle image velocimetry (SLPIV), as part of the Grand-scale Atmospheric Imaging Apparatus (GAIA). Wind tunnel data from a previous study are included for comparison (Heisel *et al.*, *J. Fluid Mech.*, vol. 887, 2020, R1). The probability density function of UMZ attributes such as their thickness, modal velocity and averaged vertical velocity are built at varying elevations and modelled using log-normal and Gaussian distributions. Inverse transform sampling of the distributions is used to generate synthetic step-like velocity profiles that are spatially and temporally uncorrelated. Results show that in the wide range of wall-normal distances and $Re_\tau$ up to $\sim O(10^6)$ investigated here, shear velocity scaling is manifested in the velocity jump across shear interfaces between adjacent UMZs, and attached eddy behaviour is observed in the linear proportionality between UMZ thickness and their wall normal location. These very same characteristics are recovered in the generated instantaneous profiles, using both fully stochastic and data-driven hybrid stochastic (DHS) models, which address, in different ways, the coupling between modal velocities and UMZ thickness. Our method provides a stochastic approach for generating an ensemble of instantaneous velocity profiles, consistent with the structural organisation of UMZs, where the ensemble reproduces the logarithmic mean velocity profile and recovers significant

† Email address for correspondence: mguala@umn.edu

portions of the Reynolds stresses and, thus, of the streamwise and vertical velocity variability.

**Key words:** turbulent boundary layers, atmospheric flows, turbulence modelling

---

## 1. Introduction

Turbulent boundary layer (TBL) flows are characterised by a high degree of complexity, which is manifested in the broad range of energetic scales that contributes to the spectrum of the streamwise velocity (Jiménez 2012; Cardesa, Vela-Martín & Jiménez 2017). These canonical flows also display a level of organisation of basic ubiquitous features, such as vortices and shear layers (Cantwell 1981; Robinson 1991; Adrian 2007; Jiménez 2018). Regions of coherent motions have been leveraged to propose simplified, low-dimensional, phenomenological models of wall turbulent flows (Perry & Chong 1982; Klewicki & Oberlack 2015; Bautista *et al.* 2019; Marusic & Monty 2019).

Laboratory investigations of the instantaneous velocity field in zero pressure gradient (ZPG)-TBL flows, under a wide range of friction Reynolds number $Re_\tau$, have shown the statistical persistence of large-scale structures in the outer region with nearly uniform streamwise velocity, denoted as uniform momentum zones (UMZs) (Meinhart & Adrian 1995; Adrian, Meinhart & Tomkins 2000; de Silva, Hutchins & Marusic 2016; Laskari *et al.* 2022). These zones are separated by internal layers of high shear (ISLs) (Eisma *et al.* 2015; de Silva *et al.* 2017; Heisel *et al.* 2021; Zheng & Anderson 2022). The ISLs are also denoted as vortical fissures (Priyadarshana *et al.* 2007), as they are densely populated by clusters of spanwise vortices (see also, Christensen & Adrian 2001; Heisel *et al.* 2018).

Due to the existence of UMZs separated by shear layers, the instantaneous velocity profiles exhibit a step-like shape (de Silva *et al.* 2016; Heisel *et al.* 2020c). A reduced-complexity, wall turbulence velocity field may thus be modelled as a sequence of discrete steps, where the ensemble average is sufficient to recover the logarithmic mean velocity profile, for both smooth- and rough-wall flows (Klewicki & Oberlack 2015; Bautista *et al.* 2019; Heisel *et al.* 2020c).

Measurements of rough wall turbulence in the atmospheric surface layer (ASL) confirmed the existence of UMZs at the field scale. Heisel *et al.* (2018) used super-large-scale particle image velocimetry (SLPIV) during snow fall events (Hong *et al.* 2014; Toloui *et al.* 2014), under near-neutral thermal stability conditions, to investigate how coherent structures grow from the laboratory to the ASL. The thickness of UMZs was observed to scale with wall-normal distance $z$, within the logarithmic region, providing a theoretical interpretation to previous laboratory observations by de Silva *et al.* (2016). Recalling the mixing length theory, $l_e = \kappa z$ in which $\kappa$ is von Kármán constant, is the size of the height-dependent eddies responsible for the turbulent momentum transfer across different flow layers (Prandtl 1925, 1932). The phenomenological interpretation of those eddies in terms of UMZs, rather than isolated swirling motions, has been recently proposed by Heisel *et al.* (2020c). A reconciling view is suggested, where the mixing length dynamics is enabled by the shear layers and vortical structures delimiting UMZs. Such description must also consider the vertical mobility of the UMZs that has not been studied in detail, so far.

Townsend (1976) originally proposed the attached eddy hypothesis (AEH), according to which energetic eddies are self-similar turbulent motions with a size proportional to the distance from the wall. The word 'attached' highlights that these structures originate at the wall and contribute to extending the shear velocity $u_\tau$ scaling throughout the outer wall region Heisel *et al.* (2020*c*). Such a scaling is manifested in the velocity difference, or jump, across the shear layer (de Silva *et al.* 2017; Heisel *et al.* 2020*c*) and in the azimuthal velocity of energetic vortex cores (Heisel *et al.* 2021). The spatial distributions of shear layers and associated UMZs are also consistent with the hairpin vortex paradigm (Adrian *et al.* 2000; Adrian 2007) and the $\Lambda$ eddy packet model (Perry & Chong 1982), supporting the idea that vortex heads tend to organise themselves along the very same shear layers separating UMZs (Heisel *et al.* 2018). Vortex organisation in the wall region is also manifested in the statistically persistent structure angle of $9°$ to $15°$ observed in the two-point correlation of the streamwise velocity fluctuations (Ganapathisubramani, Longmire & Marusic 2003; Marusic & Heuer 2007) and of the swirling strength, which is marking intense vortical flows (Christensen & Adrian 2001; Guala *et al.* 2012).

The existence of UMZs in different flow conditions and configurations confirm their ubiquity and statistical relevance: for instance, in high-Mach-number flows (Williams *et al.* 2018; Cogo *et al.* 2022), above different wall surface roughness geometries or patterns (Xu, Zhong & Zhang 2019), in adverse pressure gradient TBL flows (Bross, Fuchs & Kähler 2019), in boundary layers perturbed by surface waves (Li *et al.* 2020) and in the ASL under different thermal stability conditions (Puccioni *et al.* 2023; Salesky 2023). The UMZs and the associated shear layers can be thus framed as the representative attached eddy, coherent structure responsible for the momentum transport and the establishment of the logarithmic velocity profile over a wide range of surface roughness conditions and Reynolds numbers (Prandtl 1925; Bautista *et al.* 2019; Heisel *et al.* 2020*c*).

In this paper we address the vertical distribution of UMZs and the associated step-like instantaneous velocity profiles as the simplest representation of streamwise and vertical velocity variability in TBL flows, and we propose a hierarchy of stochastic models to reproduce such variability. Thus far, several models for generating synthetic instantaneous streamwise velocity fields have been proposed (Perry & Chong 1982; McKeon & Sharma 2010; Bautista *et al.* 2019). The most recent effort by Bautista *et al.* (2019), proposed a model based on the mean streamwise momentum equation (Klewicki *et al.* 2014) and self-similarity of turbulent structures to reconstruct the step-like instantaneous streamwise velocity profile corresponding to the presence of UMZs. The Bautista *et al.* (2019) model generates synthetic streamwise velocity profiles based on a series of UMZs, where the number of UMZs in the outer region is a function of the friction Reynolds number following de Silva *et al.* (2016), the zone thickness is imposed as a function of the wall-normal position with an assumed positively skewed Gaussian distribution, and the velocity jump between adjacent UMZs is assumed to be a constant factor of $u_\tau$. While the model of Bautista *et al.* (2019) captures first- and higher-order statistical moments of the streamwise velocity, the imposed assumptions have not been validated experimentally and the model parameters have only been evaluated for a smooth-wall direct numerical simulations (DNS) at a single $Re_\tau \sim O(10^3)$ value (Lee & Moser 2015), still relatively low as compared with ASL values $Re_\tau \sim O(10^6)$.

The focus of our work is to use experimental evidence of UMZ properties and their distribution to develop a model for the generation of step-like instantaneous velocity profiles in the wall region above rough surfaces, with no need to reach the boundary layer height. The model is formulated to mimic turbulence statistics in the logarithmic region of TBLs over a broad range of Reynolds number and surface roughness. We first leverage on the identification and statistical characterisation of UMZs geometric and kinematic attributes, acquired through extensive laboratory and field measurements, to formulate our modelling framework. For this objective, a spatiotemporally resolved velocity dataset, covering a field of view of $8 \times 9\,\text{m}^2$, was obtained in the ASL using

SLPIV, as part of the Grand-scale Atmospheric Imaging Apparatus (GAIA). The new dataset targets the roughness sublayer and the logarithmic region, consistent with the wind tunnel data from Heisel *et al.* (2020*c*), which are also included in the analysis. Two models have been proposed in this study: (i) a stochastic model, and (ii) a data-driven hybrid stochastic (DHS) model. They both rely on the stochastic generation of UMZ's thickness from cumulative distribution functions (c.d.f.s) that were parameterised using height-dependent experimental observations. The difference between the two models is in the coupling between UMZ thickness and the wall-normal and modal velocities, imposed in the generation of the synthetic profiles. Those differences percolate in the statistical description of the synthetic velocity field resulting from the ensemble of the simulated velocity profiles. Both models incorporate the vertical velocity as a UMZ attribute, which allows to extend the analysis of the synthetic velocity field to both components of the velocity variance and the Reynolds shear stress.

The long-term objective for generating stochastic modal velocity profiles is twofold: first, we aim to assess whether UMZs are responsible for a significant portion of the variability of instantaneous rough-wall turbulent flows near the surface; second, we aim to build a generalised method for generating velocity fields that could reproduce the effects of the roughness sublayer in atmospheric turbulence, interface with large eddy simulations and help develop more statistically representative wall function models

All three datasets investigated cover the lower portion of the zero-pressure gradient rough wall TBL, extending from the edge of the roughness sublayer to the logarithmic region. The bottom-up approach allows to extend the synthetic velocity profiles until needed, e.g. to potentially overlap with the lower portion of coarse computational grids employed for ASL simulations (Moeng 1984; Khanna & Brasseur 1998; Kosovic & Curry 2000; Larsson *et al.* 2016) and provide a more dynamic feedback.

The experimental datasets used in the models, the histogram-based approach for UMZ detection, and the procedure to collect UMZs at each wall normal position and estimate their statistical properties are discussed in detail in the methodology § 2. The two methods for generating synthetic step-like velocity profiles are presented in § 3. The performance of the model in reproducing the turbulence statistics of the canonical rough wall boundary layer is presented in the result § 4. The limitations and strengths of the model are in part discussed in § 5, in part summarised in the concluding § 6, with particular emphasis on the prescribed vs emerging properties of the model.

In this study, the symbol $z$ denotes the wall-normal distance, and the subscript '$i$' is utilised to denote an arbitrary elevation $z_i$ or UMZ attribute at any given location $h_{m_i}$. Each UMZ has been described by four variables: its thickness $h_m$, the mid-height elevation $z_m$, the modal velocity $u_m$ and the averaged vertical velocity $w_m$. The subscript '$m$' is used to show attributes of the UMZs, i.e. $h_m, u_m, w_m$. The superscript '+' is used for inner wall normalisation, i.e. $u^+ = u/u_\tau$. For variables, lowercase lettering indicates instantaneous value, and uppercase lettering is used for the temporal or ensemble averages. For velocity, superscript ' is used to indicate fluctuations from the mean value, i.e. $u = U + u'$. The terms 'characteristics' and 'attributes' for UMZs are used interchangeably.

## 2. Methodology

### 2.1. *Wind tunnel and ASL datasets*

Data for this investigation were in part gathered from a previously published database using PIV measurements obtained in the St. Anthony Fall Laboratory boundary layer wind tunnel with approximately ZPG and wire mesh surface roughness (Heisel *et al.* 2020*c*,*a*). Table 1 presents the experiment's parameters. The ASL measurement included here were performed at the Eolos Wind Research Field Station in Rosemount, Minnesota with SLPIV (Hong *et al.* 2014; Toloui *et al.* 2014; Heisel *et al.* 2018) and are characterised by high Reynolds number, nearly neutral thermal stability and surface roughness typical of fresh snow on flat terrain (Manes *et al.* 2008; Gromke *et al.* 2011).

| Dataset | Label | Symbol | $Re_\tau$ | $u_\tau$ (m s$^{-1}$) | $\delta$ (m) | $z_0$ (m) | Reference |
|---|---|---|---|---|---|---|---|
| Wind tunnel mesh 1 | WT (m1) | ▲ | 10 100 | 0.39 | 0.40 | $6.2 \times 10^{-4}$ | Heisel *et al.* (2020*c*) |
| Wind tunnel mesh 2 | WT (m2) | ▼ | 13 900 | 0.56 | 0.39 | $6.3 \times 10^{-4}$ | Heisel *et al.* (2020*c*) |
| Atmospheric surface layer | ASL | • | $O(10^6)$ | 0.40 | 93 | $3.3 \times 10^{-3}$ | New |

Table 1. Experimental dataset used in acquiring the profile of the statistical moments of UMZs characteristics.

### 2.1.1. *ASL measurements*

ASL measurements employed SLPIV using a near-wall configuration of the lighting system to image an approximately square $10 \times 10$ m$^2$ field of view with a $3840 \times 2160$ pixel$^2$ camera. The video was captured at a 120-Hz frame rate for 15 minutes of continuous recording, which is of the order of the averaging time used in ASL flows (Stull 1988*a*). The selected flow measurements extended from about 1 m above the surface, well above the estimated roughness sublayer $3–5k_s \sim 0.3–0.5$ m (Flack & Schultz 2014; Chung *et al.* 2021). The deployment was carried out at night on 22 February 2022, during a light snowfall. Direct sampling of snow particles provided a median (mean) particle size of $D_p = 0.65(0.82)$ mm and a mean settling velocity $W_s = 0.5$ m s$^{-1}$, subtracted from the SLPIV vertical velocity field. The estimated particle response time based on the measured settling velocity, and neglecting particle–turbulence interaction mechanisms, is estimated as $\tau_p \simeq 0.05$ s. Alternatively, using the particle response time formulation based on spherical shape and drag correction we estimate $\tau_p = \rho_p D_p^2/18\nu(1 + 0.15Re_p)^{0.687} \simeq 0.1$ s for a snow density of approximately $\rho_p = 90D_p^{-1} = 109$ kg m$^{-3}$ (Brandes *et al.* 2007). This leads to an acceptable Stokes number $St = \tau_p/\tau_f = 0.2$ for a flow time scale $\tau_f \sim 0.25$ s, corresponding to a length scale of about $l \sim \tau_f u_{rms} \sim 0.18$ m, which approximates the physical size of the $64 \times 64$ pixel$^2$ interrogation window, $2\Delta x = 0.16$ m, adopted in the SLPIV processing (with 50 % overlap). This scaling has been proposed in Hong *et al.* (2014) and Heisel *et al.* (2018) to justify the use of snow particles as acceptable tracers in the ASL, with spatial resolution targeting the organisation of large-scale flow features, such as UMZ and shear layers. Detailed analysis by Heisel *et al.* (2021) on a similar SLPIV dataset showed that the shear velocity scaling of both the interface velocity jump, between adjacent UMZs, and of the azimuthal velocity of identified energetic vortices is preserved. This suggests that the response of snow particles to the turbulent flow scales resolved by SLPIV is adequate for the purposes of the present analysis. We acknowledge that the inertial snow properties leading to the scale-dependent Stokes number, and the nominal light sheet thickness of about 0.3 m represent the current limiting factors preventing a further increase in resolution. Based on velocity and temperature measurements from a colocated sonic anemometer, the thermal stability condition was classified as near-neutral with $|z/L| < 0.008$ (Iungo *et al.* 2023). The Monin Obhukov length, $L$, is in line with the conditions explored by Toloui *et al.* (2014) and Heisel *et al.* (2018), during similar snowfall events. By fitting the mean velocity profile with logarithmic law, the shear velocity was estimated as $u_\tau = 0.40\ \text{m s}^{-1}$ and the Coriolis parameter $f_c$ was calculated based on the latitude of the field site. The ASL depth $\delta = 0.03 u_\tau / f_c = 93\ \text{m}$ was estimated assuming 10 % of the atmospheric boundary layer depth, weakly stable thermal conditions ($z/L = 0.008$) and geostrophic equilibrium between Coriolis acceleration and the pressure gradient (Stull 1988*b*; Zilitinkevich & Chalikov 1968). We acknowledge that there is significant uncertainty in the estimate of the aerodynamic roughness length $z_0$ and shear velocity, likely due to contamination from the roughness sublayer and by unsteadiness and variability of the inflow conditions in the field, including those driven by weak thermal stability. Comparison with the colocated sonic anemometer suggests an uncertainty of 0.05 m s$^{-1}$ in the shear velocity and a range of $z_0$ values from 0.001 to 0.003 m. Figure 1 provides a picture of the deployment and a sample of the SLPIV fluctuating velocity field. Further details of the deployment are included in Iungo *et al.* (2023).

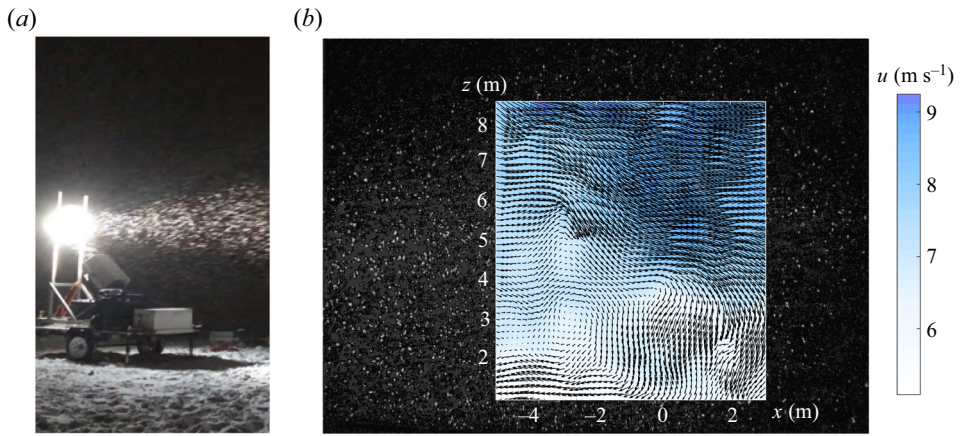

Figure 1. ASL experimental set-up: (*a*) photo of the vertical light–mirror configuration for near-surface illumination; (*b*) sample recorded image and instantaneous fluctuating velocity field superimposed on the streamwise velocity contour (a constant $u_0 = 6.7\ \text{m s}^{-1}$ advection velocity is subtracted).

### 2.2. *UMZ detection*

UMZs have been identified in each dataset by detecting local maxima in the histograms of the instantaneous streamwise velocity $u$, following the step-by-step procedure of Heisel *et al.* (2020*c*). This approach, introduced by Adrian *et al.* (2000), is also known as the histogram-based approach for UMZ recognition. Alternative methods have been proposed recently by others (Fan *et al.* 2019; Yao *et al.* 2019; Younes *et al.* 2021).

We briefly present here the methodology and the associated parameters. Every PIV snapshot in the wind tunnel datasets is subdivided into three subsamples to contain roughly 7000 vectors and cover the roughness sublayer and the entire logarithmic region. The instantaneous streamwise velocity field shown in figure 2(*a*) is the foundation upon which the velocity histogram is calculated. The sample size was chosen to correctly identify physically relevant histogram peaks, defining the modal velocities $u_m$ and associate them with coherent flow regions. A lower sample size would lead to more peaks in the histogram due to random turbulent fluctuations, whereas a larger sample size would smooth the distribution due to overlapping UMZ travelling at different modal velocities. We direct the reader to Heisel *et al.* (2018) and the discussion in de Silva *et al.* (2016) and Heisel *et al.* (2020*c*) for a more systematic analysis of the sample size effect on UMZ detection. In the following, we focus on the analysis of the new ASL dataset and on the specific UMZ detection parameters as compared with the wind tunnel datasets. First, to maintain in the ASL dataset a number of vectors per sample consistent with the wind tunnel datasets and improve the statistical convergence of the histogram, three successive PIV snapshots ($t - \Delta t$, $t$, $t + \Delta t$) were taken into account in the computation of the histogram of the streamwise velocity at each time $t$. In addition to the vector sample size, the velocity histogram is affected by the streamwise extent of the measurement domain $\mathcal{L}_x$. If $\mathcal{L}_x$ is substantially larger than the size of the expected UMZ, then numerous regions with coherent (but distinct) velocity would contribute to the histogram, possibly smearing it and concealing the smallest local peaks. Heisel *et al.* (2020*c*) computed the average UMZ thickness $H_m$ in the logarithmic region and showed that $\mathcal{L}_x$ acts as a low-pass filter for the velocity histogram. To improve UMZ detection, it is thus desirable that the variability in the streamwise velocity histogram emerges from vertically stacked flow regions travelling at different modal velocities and resembling a step-like instantaneous velocity profile. To facilitate UMZ extraction, in the ASL the field of view was defined by a relatively small streamwise extent, $\mathcal{L}_x = 8$ m, while covering a region of intense shear extending up to $\mathcal{L}_x/\delta \approx 0.1$, which is comparable with the wind tunnel datasets.

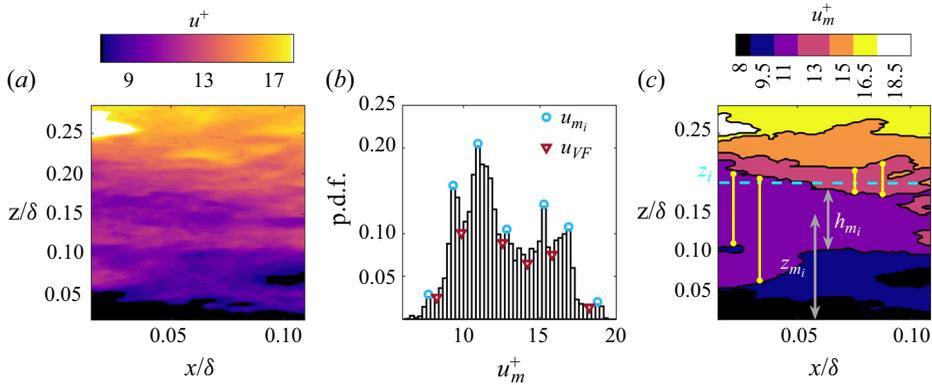

Figure 2. Example of UMZ detection methodology from experiment WT (m1) in table 1. (*a*) Instantaneous streamwise velocity field. (*b*) Histogram of the velocity vectors in (*a*) with detected modal velocities $u_{m_i}$ as blue circles and shear velocities $u_{VF}$ as inverted red triangles. (*c*) Thickness $h_{m_i}$, mid-height elevation $z_{m_i}$ and normalised modal velocities $u_m^+$ for the detected UMZs. Yellow vertical lines mark sampled thicknesses $h_{m_i}$ of the UMZs intersecting the reference height $z_i$ (dashed horizontal line) accounted for in height-specific statistics.

The ratio of the streamwise extent of wall-attached structures to their wall-normal position is expected to be in the 10–15 range (Baars, Hutchins & Marusic 2017). Accordingly, the typical length of UMZs at our lowest observed wall-normal position in the wind tunnel, $z = 0.025\delta$, is about $0.2\delta$. The chosen value $\mathcal{L}_x = 0.1\delta$ is therefore adequate, for both the wind tunnel and the ASL, to capture relatively small UMZs located near the edge of the roughness sublayer, as well as progressively larger UMZs growing in the logarithmic region.

In addition to the streamwise length $\mathcal{L}_x$, the number of vectors per sample, and the vertical extent of the shear region covered, also the bin width influences the peaks of the velocity histogram. The normalised bin width for the velocity histogram in figure 2(*b*) is set to $0.3u_\tau$, which is close to the $0.5u_\tau$ reported by Laskari *et al.* (2018) and slightly larger than the SLPIV velocity uncertainty $\simeq 0.1 \ \mathrm{m\,s^{-1}}$ reported by Heisel *et al.* (2018). It has been demonstrated that UMZs are separated by thin layers characterised by strong velocity gradients, where shear is concentrated (Laskari *et al.* 2018) and spanwise vortices are likely to reside (Heisel *et al.* 2018). According to de Silva *et al.* (2017), the velocity jump between two neighbouring UMZs, estimated as the difference between the respective modal velocities, is in the range of one to two times the friction velocity $u_\tau$. Therefore, the $0.3u_\tau$ discretisation of the velocity distribution is expected to be sufficient to detect distinct modal velocities in each histogram. The local peaks which are illustrated by blue circles in figure 2(*b*), are found using a peak detection algorithm based on a pretested set of parameters: the minimum distance between two peaks ($0.5u_\tau$), the minimum peak prominence, i.e. the height difference between the peak and its adjacent minima ($2 \times 10^{-4}$ for a normalised, unit area, histogram). These parameters are comparable with previous values (Laskari *et al.* 2018; Heisel *et al.* 2020*c*). A sample outcome of the UMZ detection algorithm is presented in figure 2(*c*), with the modal velocity displayed as blue circles on the histogram in figure 2(*b*). The local minima between the detected peaks, identified by red inverted triangles in figure 2(*b*), are used to identify the shear layers, or interfaces, between UMZs.

In figure 2(*c*), black curves mark the corresponding interfaces based on isocontours of the histogram local velocity minima. At each streamwise location $x$, the UMZ thickness $h_{m_i}$ is the vertical distance between UMZ interfaces. The thickness and its mid-height elevation $z_{m_i}$ are depicted as grey double arrow in figure 2(*c*).

In order to quantify the contribution of UMZs to the Reynolds shear stress and the corresponding partitioning into sweeps ($u' > 0$, $w' < 0$) and ejections ($u' < 0$, $w' > 0$), the vertical velocity $w_{m_i}$ is taken into account as a UMZ attribute. Specifically, it is estimated at each streamwise position $x$ as the spatially average vertical velocity along the thickness $h_{m_i}$. Lastly, the thickness of every detected UMZ is compared with the Taylor micro-scale estimated at the mid-height $z_{m_i}$ to filter out possible outliers, such as travelling vortices or shear layers (Heisel *et al.* 2021). The Taylor micro-scale, $\lambda = (15\nu/\varepsilon)^{1/2}$ is estimated through hot-wire measurements for the wind tunnels and sonic anemometer data in the ASL; the turbulent kinetic energy dissipation rate $\varepsilon$ is estimated using the second-order structure function of the streamwise velocity following Saddoughi & Veeravalli (1994). The full database comprises $O(10^6)$ UMZs. Zones reaching the top and bottom of the PIV field are excluded from the analysis to avoid statistical bias (e.g. showing a prevalence of thin UMZs near the top of the PIV domain). For the wind tunnel datasets, we extended the vertical range of UMZ detection closer to the wall, within the roughness sublayer, as compared with the $z/\delta > 0.05$ lower limit adopted in Heisel *et al.* (2020*c*). In this experimental set-up the mean velocity remains well approximated by the logarithmic law within the roughness sublayer, and the roughness effects are more apparent in higher-order statistics (Heisel *et al.* 2020).

### 2.3. *UMZs height-dependent collection*

Following the identification of UMZs at each column of each PIV frame, we present now the method to combine the descriptive attributes $h_m$, $u_m$, $w_m$ at each generic elevation $z$. This is critical to formulate height-dependent c.d.f.s, which are at the core of the UMZ generation stochastic process.

*Stochastic modelling of rough-wall turbulence*

Consistent with the definition of UMZ thickness, we define the elevation range, centred around the mid-height position, extending from $z_m - h_m/2$ to $z_m + h_m/2$ (Heisel *et al.* 2020c). Figure 2(c) provides an example of how the statistical distribution of UMZ thickness and mid-height is extracted. At each given $z_i$, illustrated by the blue dashed line, we collect all UMZs that intersect $z_i$, marked in yellow and delimited by double circles. In this way, any UMZ property can be conditionally averaged on the distance from the wall $z = z_i$, thus accounting also for large UMZ close to the surface, where $z_m > z_i$. The distribution of $h_m$ for three wall-normal positions is shown in figure 3(a–c) for the three datasets, and classified as log-normal. The most probable UMZ thickness is observed to increase with $z$, consistent with the behaviour of attached eddies. The normalised UMZ modal velocity $u_m^+$ and the associated vertical averaged velocity $w_m^+$ are plotted in figure 3(d–i) and are approximately Gaussian. Due to the mean shear, the UMZ streamwise modal velocity increases with wall normal distance $z$, whereas the mean vertical velocity stays around zero.

The persistent wall-normal dependent behaviour of UMZs in the log region clearly emerges for all datasets in the joint probability of $h_m/\delta$ and $z/\delta$, in figure 4. As discussed in Heisel *et al.* (2020c), the size of statistically dominant eddies appear consistent with the thickness of the UMZ regions, $l_e \sim H_m(z) = 0.75z$, and scales linearly with $z$ as originally proposed in the mixing length theory (Prandtl 1932). Note that a similar result, $l_e = 0.62z$, based on UMZ modelling was proposed by Bautista *et al.* (2019) for smooth wall boundary layers. We stress that figure 4 offers a robust scaling law to predict the average UMZ thickness at different heights, whereas figures 3(a)–3(c) describe their distribution and variability.

### 2.4. *Modelling parameters for a statistical description of UMZs*

The estimated distributions of UMZ properties can be described by specific families of mathematical functions. We use the log-normal distribution for the thickness $h_m$ and the Gaussian distribution for $u_m$ and $w_m$. These functions require two $z$-specific modelling parameters that depend on the mean $\mu$ and the standard deviation $\sigma$ of the UMZ attributes collected at each elevation $z$.

In particular, the log-normal distribution of $h_m$ requires as modelling parameters the mean $\hat{\mu}_{H_m}$ and the standard deviation $\hat{\sigma}_{H_m}$ of the logarithm of the thickness of all identified UMZs collected at each wall-normal position. These are dimensionless and defined as

$$\hat{\mu}_{H_m} = \frac{1}{n}\Sigma_k \ln\left(\frac{h_{m_k}}{1}\right) \quad \text{and} \quad \hat{\sigma}_{H_m} = \sqrt{\frac{1}{n-1}\Sigma_k \left(\ln\left(\frac{h_{m_k}}{1}\right) - \hat{\mu}_{H_m}\right)^2}, \quad (2.1a,b)$$

where $n$ is the number of the UMZ thickness measurements collected at $z$ and a nominal 1(m) width in the denominator of fractions is used for non-dimensionalisation (see the discussion section for a more generalised normalisation). Please note that $\hat{\sigma}_{H_m} = \sqrt{ln(1 + \sigma_{h_m}^2/\mu_{h_m}^2)}$, implying that while the mean and variability of $h_m$ increase at the field scale, the parameter $\hat{\sigma}_{H_m}$ remains comparable with the wind tunnel data. The mean and standard deviation of the modal $u_m$ and vertical $w_m$ velocities are the two parameters of the corresponding Gaussian distributions calculated for the collected UMZs at each wall-normal position. The statistics of all UMZ attributes are illustrated in figure 5 as a function of $z$, for the three datasets. At this stage, we do not aim to formulate predictive models for these curves, but we need a mathematical expression, for each UMZ attributes, to build the c.d.f. at any arbitrary wall-normal location and reproduce the variability of the experimental observations.

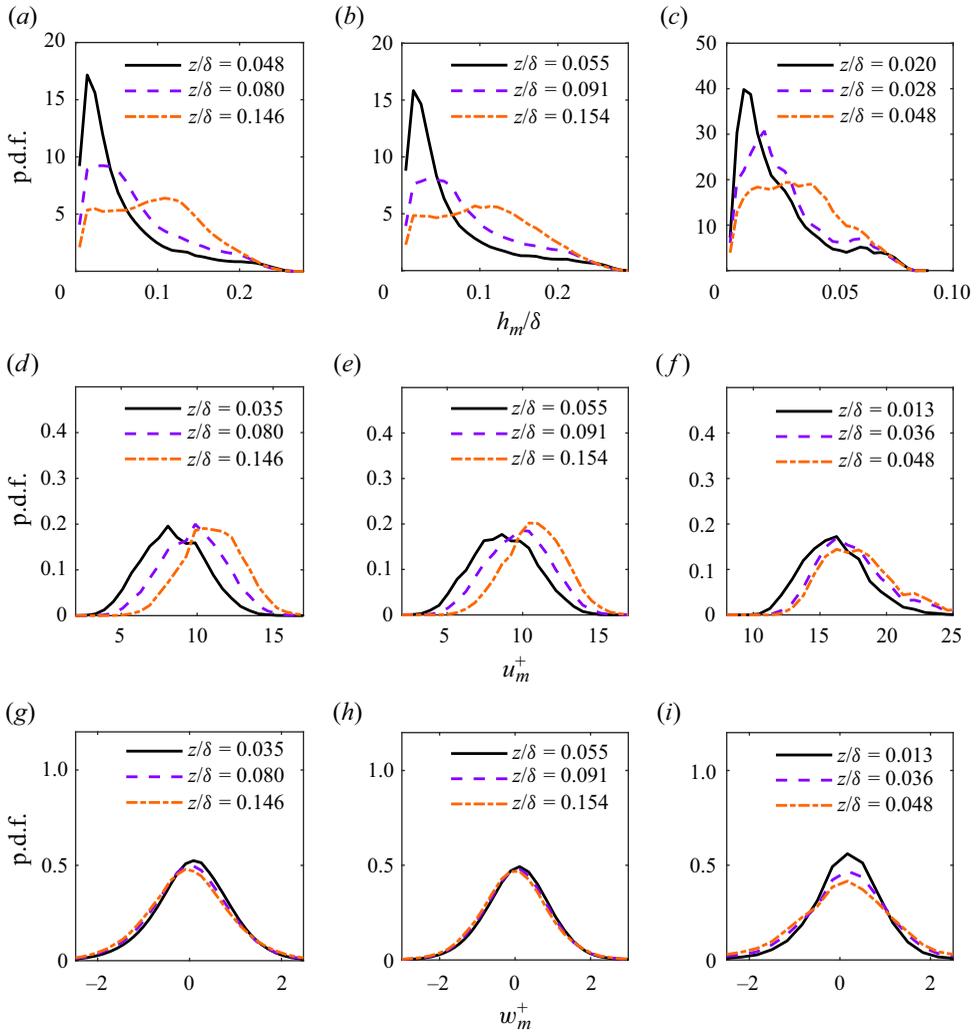

Figure 3. Probability density function (p.d.f.) of $h_m/\delta(z)$, $u_m^+(z)$, and $w_m^+(z)$ at three wall-normal position for the three datasets: $(a,d,g)$ from wind tunnel WT (m1), $(b,e,h)$ from wind tunnel WT (m2) and $(c,f,i)$ from the ASL.

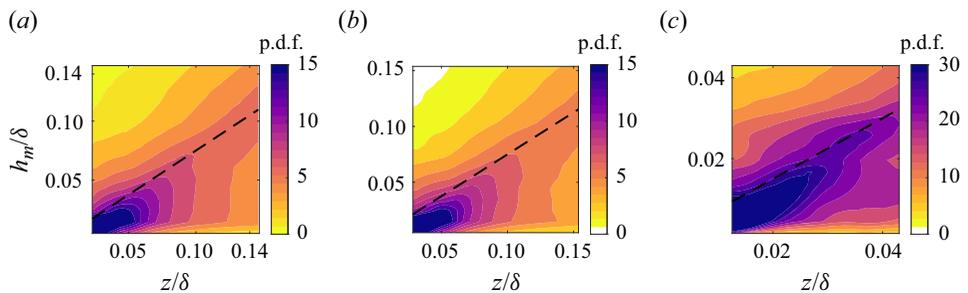

Figure 4. Normalised joint p.d.f. of $h_m/\delta$ and $z/\delta$: $(a)$ wind tunnel dataset WT (m1) and $(b)$ wind tunnel dataset WT (m2); $(c)$ ASL dataset. Dashed lines indicate $H_m(z) = 0.75z$, for reference.

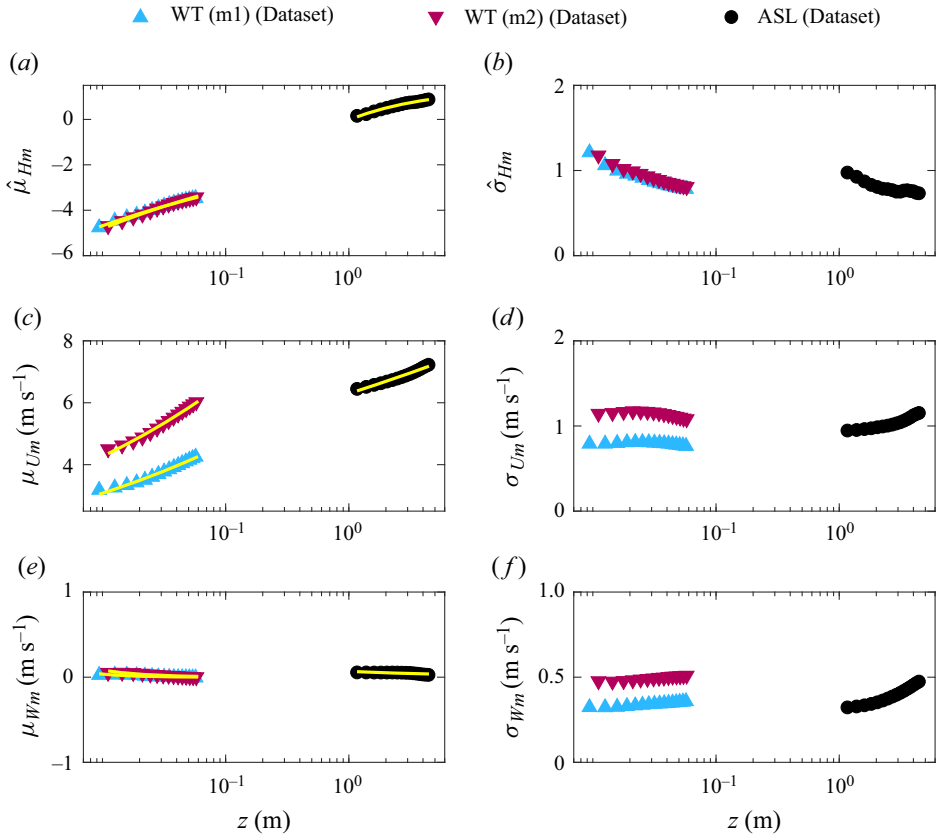

Figure 5. Required parameters to reproduce p.d.f. and c.d.f. of three UMZ attributes at different wall-normal positions, for the three datasets: (a) mean of the logarithm of the extracted thicknesses $\hat{\mu}_{H_m} = \mu(\log(h_m))$; (b) standard deviation of the logarithm of the extracted thicknesses $\hat{\sigma}_{H_m} = \sigma(\log(h_m))$; (c) mean of the modal velocity distribution $\mu_{U_m}$; (d) modal velocity distribution's standard deviation $\sigma_{U_m}$; (e) mean of the distribution of the wall-normal velocity $\sigma_{W_m}$; (f) standard deviation of wall-normal velocity $\sigma_{W_m}$. Yellow curves in panels (a,c) represent the fitted power law and logarithmic functions, respectively, for the first statistical moment of the corresponding variables.

Note that the specific method of collecting UMZs, at each given wall-normal position, plays a role in the observed log-normal distribution of UMZ thickness (figure 3a–c). We acknowledge that the vertical profile of the mean of the thickness logarithms (figure 5a) does not preserve the linear trend proportional to the wall distance $z$ characteristic of the AEH and displayed in the joint p.d.f. in figure 4. This is a contamination effect that would not exist if we only collected UMZ based on their mid-height position $z_m$. However, such a choice would introduce a significant bias in the data because near the wall or the upper edge of the observation domain, only very thin zones would be extracted. Owing to the above reasons, we preferred to collect unbiased intersecting UMZs at any given height $z_i$, as described previously, and employ a generic power-law function to model $\hat{\mu}_{H_m}(z)$ (yellow line), as shown in figure 5(a). We confirm in the results section that the statistics computed on the generated UMZ profiles, including the joint p.d.f. $(h_m, z)$, accurately reproduce the experimental trends.

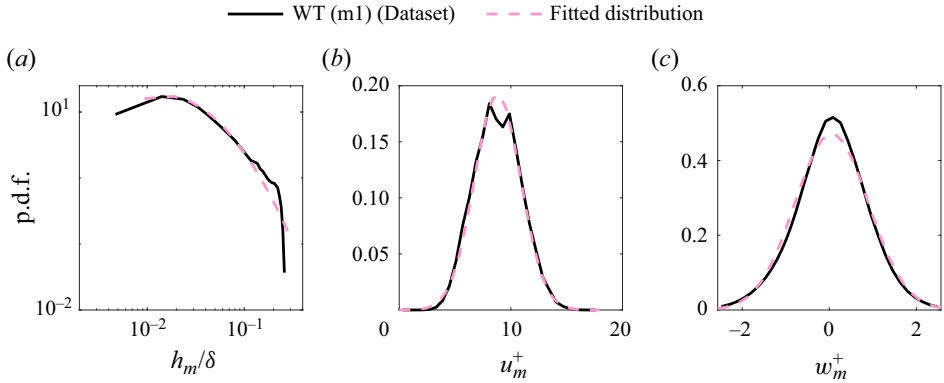

Figure 6. Probability density function of normalised ($a$) UMZ's thickness $h_m/\delta$, ($b$) UMZ's modal velocity $u_m^+$, ($c$) UMZ's vertical velocity $w_m^+$, estimated from the wind tunnel WT(m1) dataset, at wall-normal position $z/\delta = 0.048$ (solid line) and reconstructed from the parameters in figure 5 (dotted line).

The mean of the modal velocities at various elevations are expected to converge to the mean streamwise velocity profile (Heisel *et al.* 2020$c$), hence we choose a logarithmic function for $\mu_{U_m}(z)$ (yellow line in figure 5$c$). The mean vertical velocity $\mu_{W_m}(z)$ is approximately zero for all elevations, as expected in nearly zero-pressure gradient boundary layer flows (figure 5$e$). Regarding the second parameters of the above distributions, e.g. the variances, $\hat{\sigma}_{H_m}, \sigma_{U_m}, \sigma_{W_m}$, we do not have any theoretical argument to support specific functions of wall-normal distance, within the explored range of heights, and we rely on the interpolation between experimental observations. A preliminary, approximate formulation of normalised statistical moments of UMZ attributes, in the explored ranges of $Re_\tau$ and $z/\delta$, is presented in the discussion section. Figures 6($a$)–6($c$) illustrate an example of experimentally derived and reconstructed distributions for UMZ thickness, modal velocity and vertical velocity, respectively, at wall-normal position $z/\delta = 0.048$.

## 3. Stochastic generation of step-like instantaneous velocity profiles

Two methods are used to generate synthetic instantaneous streamwise velocity profiles from the estimated distributions. Both methods start from the first point above the surface, where the c.d.f. of the step height is reconstructed from the two height-dependent parameters $\hat{\mu}_{H_m}$ and $\hat{\sigma}_{H_m}$. Figure 7($a$) illustrates an example of the reconstructed c.d.f. for UMZ thickness at a given wall-normal position $z$. A random number between 0 and 1 is sampled from a uniformly distributed population [$r_{h_{m_i}} \in (0\ 1)$]. The first UMZ thickness, or step height, $h_m$, is produced by inverting the c.d.f. of the log-normal distribution defined by the near-wall parameters extracted from the modal velocity field. This technique is known as inverse transform sampling and has been used in different research fields (Foufoula-Georgiou & Stark 2010; Fan *et al.* 2016; Heisel 2022). In particular, the corresponding UMZ thickness is estimated as $h_{m_i} = \exp(\hat{\mu}_{H_m}(z_i) + \hat{\sigma}_{H_m}(z_i)\sqrt{2}\ \mathrm{erfinv}(2r_{h_{m_i}} - 1))$, where erfinv is the inverse of the error function.

To estimate the modal and vertical velocity associated with the generated UMZ thickness, two approaches are suggested, marking the core differences between the two proposed models. Through the first approach, here defined simply as 'stochastic' and based again on inverse transform sampling, the c.d.f. of the modal velocity is reconstructed from the estimated parameters at the mid-height elevation of generated UMZ, i.e. $z_{m_i} = z_i + h_{m_i}/2$. Then, the reconstructed c.d.f. is inverted and a second random number $r_{u_{m_i}}$ sampled from a uniform distribution is used to determine the modal velocity of the generated UMZ. As discussed in § 5, a detailed analysis on the correlation between extracted $h_m$ and $u_m$ was performed to assess the dependency between $r_{u_{m_i}}$ and $r_{h_{m_i}}$. Eventually, two uncorrelated random numbers were preferred, also to avoid a systematic contamination effect on the near-wall modal velocity by the statistical weight of thicker zones, contributing more to the height-specific ensemble average statistics.

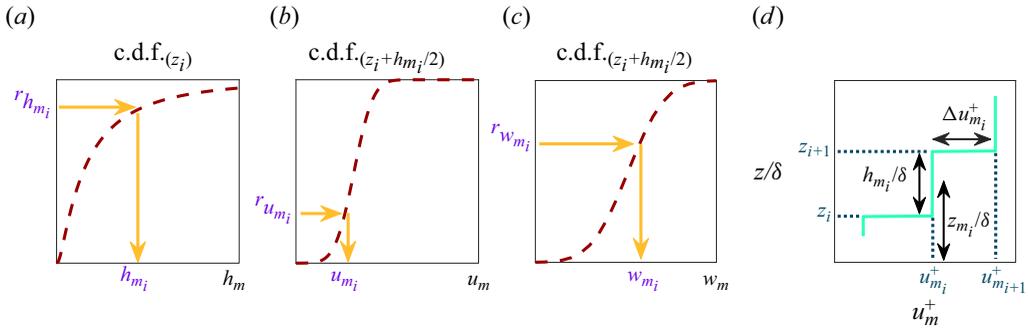

Figure 7. Stochastic model procedure for generating UMZ attributes by inverse transform sampling of the c.d.f. (red dashed line) of (*a*) UMZ thickness $h_m$, (*b*) UMZ modal velocity $u_m$ and (*c*) UMZ vertical velocity $w_m$; (*d*) resulting instantaneous velocity step profile.

For the vertical velocity, in the stochastic method we could not generate a third independent random number $r_{w_{m_i}}$ uncorrelated with $r_{u_{m_i}}$ because it would lead to an unrealistic $u'_m$–$w'_m$ distribution and to a precisely zero contribution to the Reynolds shear stress. Therefore, we generated $r_{w_m}$ and $r_{u_m}$ from a Gaussian copula with linear correlation parameter equal to $\rho_{r_{u_m}, r_{w_m}}$, ensuring a desired Pearson correlation coefficient and uniform distribution of both numbers in the $\in (0\ 1)$ range (Joe 1997; Nelsen 2007). The value of the correlation is estimated using the corresponding datasets ($\rho_{r_{u_m}, r_{w_m}} = -0.4$, for wind tunnel, and $\rho_{r_{u_m}, r_{w_m}} = -0.22$ for the ASL dataset, both fairly independent of $z$ within the explored range). It should be noted that for the reconstruction of the c.d.f. of the vertical velocity, the statistical parameters of mid-height elevation are utilised, i.e. $\mu_{W_m}(z_{m_i})$, $\sigma_{W_m}(z_{m_i})$. Figure 7(*b*,*c*) show an example of the height-dependent, reconstructed c.d.f. for the UMZ modal and vertical velocity. These UMZ attributes were determined using the inverse equation of the c.d.f. of the normal distribution for each sampled random number: $u_{m_i} = \mu_{U_m}(z_{m_i}) + \sigma_{U_m}(z_{m_i})\sqrt{2}\,\mathrm{erfinv}(2r_{u_{m_i}} - 1)$, $w_{m_i} = \mu_{W_m}(z_{m_i}) + \sigma_{W_m}(z_{m_i})\sqrt{2}\,\mathrm{erfinv}(2r_{w_{m_i}} - 1)$. The modal velocity profile shown in figure 7(*d*) is just one of the many step profiles generated by the model.

In the second model, the modal and vertical velocities are evaluated using a data-driven approach based on the actual measured UMZ attributes. We define this method as DHS. The thickness is determined using the above stochastic modelling, which is illustrated in figure 8(*a*), and the wall-normal position ($z_i$) is used to calculate mid-height elevation of UMZ ($z_{m_i} = z_i + h_{m_i}/2$).

Using the two generated UMZ attributes, $h_{m_i}$ and $z_{m_i}$, as our foundation, we search the physical dataset to find the measurements that most closely match the generated values. For each UMZ, we employed a *n*-nearest-neighbour approach based on Euclidean distance, between the generated pair $h_{m_i}$, $z_{m_i}$ and the data pairs ($h_{m_j}$, $z_{m_j}$, for $j = 1, \ldots, n$), in order to assign the values of vertical and streamwise modal velocities $u_{m_i}$, $w_{m_i}$. Figure 8(*b*) shows how to carry out this process. Following the selection of the *n*-nearest neighbours, each neighbour is assigned a weight $C_j = 1/D_j^2$ in which $D_j$ is the Euclidean distance from the target. The weighted average of the *n*-nearest neighbour provides the value for the modal and vertical velocity of the corresponding UMZ ($u_{m_i} = \sum_{j=1}^{n} C_j u_{m_j} / \sum_{j=1}^{n} C_j$, $w_{m_i} = \sum_{j=1}^{n} C_j w_{m_j} / \sum_{j=1}^{n} C_j$). This method is considered hybrid stochastic modelling because it first utilises stochastic modelling for determining UMZ thickness, and then searches, within the actual dataset, the closest UMZs that embrace the same thickness and mid-height elevation to determine the corresponding modal and vertical velocity. We explore in Appendix A the adoption of different numbers of neighbours chosen for computing the modal and vertical velocity. Eventually, we selected $K_{u_m} = 1$ and $K_{w_m} = 1$ to ensure the correct reproduction of the streamwise $\overline{u'_m u'_m}$ and vertical velocity variances $\overline{w'_m w'_m}$.

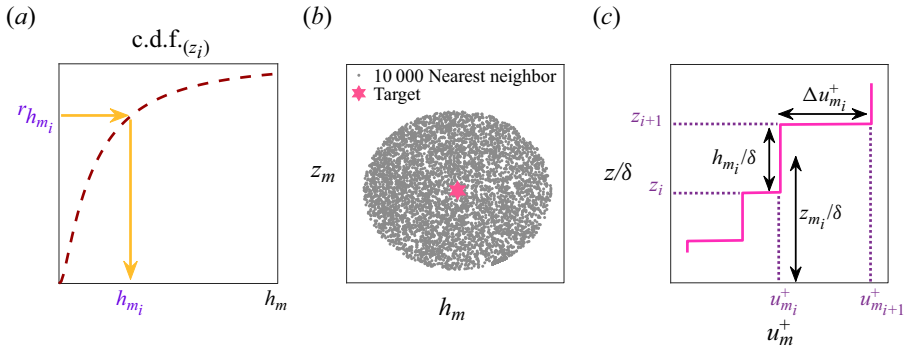

Figure 8. DHS method: (*a*) $h_m$ is generated by inverse transform sampling; (*b*) modal and vertical velocities are assigned by the nearest-neighbour algorithm given UMZ thickness $h_m$ and vertical location $z_m$; (*c*) resulting step velocity profile.

With the generated UMZ thickness, the new wall-normal position is computed as $z_i = z_{i-1} + h_{m-1}$. The previously specified steps are carried out again, iterating this process until the end of the log region, or any height below, as demonstrated in figure 8(*c*). Please note that the bottom-up generation procedure uses *z*-specific, hence local, attributes to generate the next UMZ step. The local c.d.f. of $u_m$ and $h_m$ however reflects the way we collect zones contributing at each height $z = z_i$ within the identified UMZ thickness. Therefore, in the bottom-up approach the synthetic step-like generation is weakly, though systematically, contaminated by UMZs that tend to populate flow regions farther from the wall. This is compensated for by the newly generated UMZ which extends vertically from the step bottom height ($z_i$), compensating for the above potential bias. The process of generating velocity profiles is repeated until the emerging ensemble statistics converge, as discussed in the following.

## 4. Results

The generation of a large number of instantaneous velocity profiles featuring UMZs with different attributes (thickness, modal and vertical velocities) allows us to build a multidimensional dataset that can be interrogated as an ensemble of spatiotemporally uncorrelated realisations. The convergence of the statistics derived by such an ensemble, for the ASL case, is assessed in figure 9 where the mean velocity at mid-elevation $z/\delta = 0.04$ is computed over an increasing number of samples $N_s$, using the two stochastic-based methods described previously. The convergence of the synthetic results are compared with the convergence of experimental data. Note that by taking temporally correlated PIV vector fields at $120\,\mathrm{Hz}$, $U(t)$ the number of samples increases significantly. However, the convergence of the mean velocity still suffers from large-scale unsteady motions that are known to characterise the ASL (Hutchins & Marusic 2007; Guala, Metzger & McKeon 2011; Hutchins *et al.* 2012; Puccioni *et al.* 2023), and it is thus intrinsically related to the choice of the averaging time and possible filtering strategies separating the turbulence from a slowly varying mean flow. When downsampled in time by approximately the integral time scale, the statistically independent experimental profiles $U(i)$ better resemble the sequence of the numerically generated profiles; their convergence is seemingly achieved after about $O(10^2)$ realisations, comparably with the stochastic procedures.

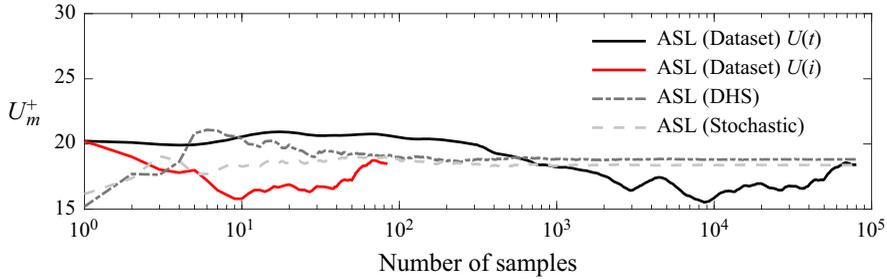

Figure 9. Convergence of the mean modal velocity $U_m$ using the stochastic and DHS approaches, for a point at $z/\delta = 0.04$ in the ASL database, as compared with the mean streamwise velocity profile from experimental data (black solid line) sampled, as acquired, at $120\,\mathrm{Hz}$ $U(t)$, or downsampled by the local integral time scale $U(i)$.

Figure 10 illustrates the wall-normal profiles of the averaged velocity as obtained by stochastically generated and experimental datasets. The theoretical logarithmic law for rough wall TBLs $U/u_\tau = (1/\kappa)\ln(z/z_0)$ is recovered. Here $\kappa = 0.39$ is the von Kármán constant, whereas $z_0$ is the aerodynamic roughness length. The only marked deviations are due to the adoption of the modal velocities in the closest UMZs to the wall, which could not capture the near-surface velocity gradient of the measured dataset. The relatively small coherent features nearest to the surface do not manifest as distinct histogram peaks, either due to limitations in the detection methodology or the lack of UMZs in that region. As a result, the roughness sublayer is often included in a relatively faster UMZ centred farther from the surface, leading to overestimated near-surface velocities in the model.

Note that the non-ideal representation of UMZs very close to the surface enhances the discrepancy between the DHS and the stochastic method: the stochastically generated modal velocity are anchored to the mean modal velocity resulting from UMZ averaging at each $z_i$ and converging to the local mean PIV velocity; the DHS model, instead, converges to the ensemble average of the modal velocity field, which is overestimated close to the wall. In the ASL, near-wall UMZs identification is more challenging and, thus, the DHS methods leads to a larger deviation from the logarithmic velocity profile. The better performance of the stochastic model in reproducing the mean velocity profile thus relies on the robustness of the distribution parameters informed by the log law, as opposed to the difficulty to obtain a flawless height-dependent joint distribution of $h_m$ and $u_m$. In other

高

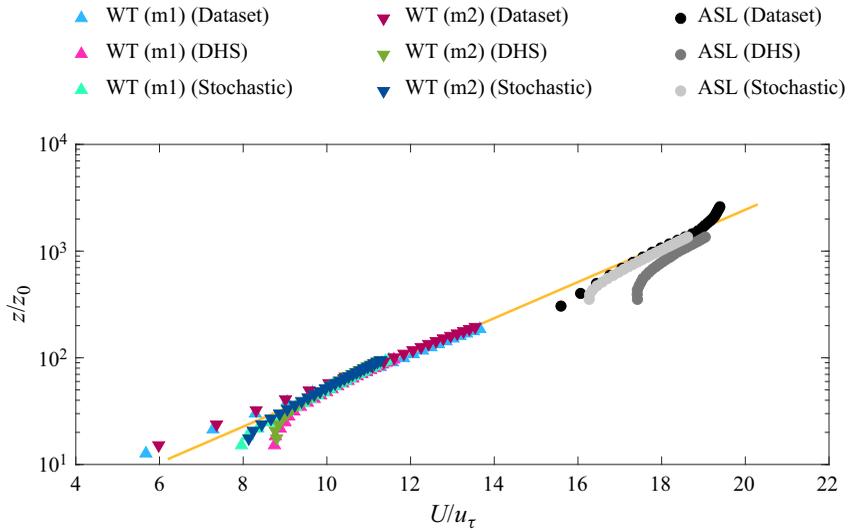

Figure 10. Normalised mean streamwise velocity profile from the experimental data and from the ensemble of profiles generated by the stochastic and DHS methods. The yellow line marks the theoretical logarithmic law for rough walls.

words, the DHS approach is penalised by the shortcomings of the UMZ detection and identification near the wall.

The height-dependent velocity variances can also be computed from the ensemble of generated profiles. Results are plotted in figure 11 where the streamwise and vertical velocity variances of the generated profiles, $\overline{u'_m u'_m}$ and $\overline{w'_m w'_m}$, respectively, are normalised by the variances measured directly from the PIV vector fields. The plots show that despite differences in Reynolds number and aerodynamic roughness, the synthetic modal velocity profiles from both models approximately reproduce the total streamwise and vertical velocity variances. The overestimate near the wall is consistent with a similar comparison shown in Heisel *et al.* (2018) where UMZ were extracted in the spatio($z$)–temporal($t$) domain sampled by PIV measurements in the ASL. Note that: (i) vortices and high-frequency fluctuations are not featured in the stochastic generated step-like velocity profiles; (ii) the DHS modelled variance depends on the number of neighbours (discussed in Appendix A). However, the majority of the variance is dictated by the presence of large- and very-large-scale motions and the associated scale interaction mechanisms (see, e.g., Hutchins & Marusic 2007; Guala *et al.* 2011; Peruzzi *et al.* 2020; Jacobi *et al.* 2021). The signature of these features is reflected in the UMZ statistical behaviour and in the estimated probability distribution of UMZ attributes. Thus, the synthetic UMZs represent the large-scale portion of the variance. While the small-scale component of the fluctuations is missing from the synthetic profiles, its absence is more appreciable for $\overline{w'_m w'_m}$ in figure 11(*b*) than for $\overline{u'_m u'_m}$ in figure 11(*a*).

The fluctuations of the streamwise modal and vertical velocities allow for the calculation of the Reynolds shear stress contribution of each UMZ. Wall-normal profiles of $-\overline{u'_m w'_m}$ are divided by $u_\tau^2$ ($u_\tau^2 \simeq -\overline{u'w'}$) and are shown in figure 12. The prescribed correlation between the random numbers used for the determination of modal and vertical velocity ($r_{u_{m_i}}, r_{w_{m_i}}$) allow the reproduction of Reynolds shear stresses by the stochastic method. For the DHS model, both modal velocity components are assigned selecting the zone most closely identified with the generated one (for $K = 1$). The limited spatial resolution in the ASL dataset and the relatively large value of the Stokes number, marking a compromised ability of a snowflake to closely follow fluid parcel trajectories (Samimy & Lele 1991), are both responsible for the underestimate of the Reynolds shear stress and for the uncertainty in the friction velocity. This limitation, which appears less stringent for the velocity variances, affects the comparison between some results emerging from the modal velocity field and the generated profiles, with those computed on the instantaneous velocities from the high-Reynolds-number dataset. The most significant effect is in the underestimate of the ratio $\rho_{u'_m w'_m} = \overline{u'_m w'_m} / \sigma_{u'_m} \sigma_{w'_m}$ which is the correlation coefficient between $u'_m$ and $w'_m$. This ratio reduces from the canonical value of $-0.4$ (Sillero, Jiménez & Moser 2013; Squire *et al.* 2016; Heisel *et al.* 2020) estimated in our wind tunnel to $-0.22$ for the ASL dataset. We also acknowledge that factors such as heterogeneous surface roughness, large-scale unsteadiness of the ASL, thermal stability affects, may cause some departures from the canonical behaviour, as the observed increase of turbulent intensity with elevation, or the uncertainty in the estimate of the friction velocity and in the normalisation of the Reynolds stresses.

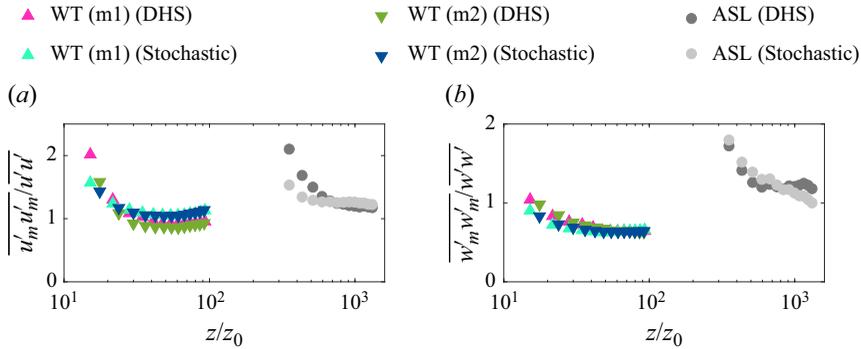

Figure 11. Profiles of (*a*) the generated modal velocity variance $\overline{u'_m u'_m}$ normalised by the PIV streamwise velocity variance $\overline{u'u'}$ and (*b*) generated wall-normal UMZ velocity variance $\overline{w'_m w'_m}$ normalised by the PIV vertical velocity variance $\overline{w'w'}$. Results are provided for the three datasets and the two methods.

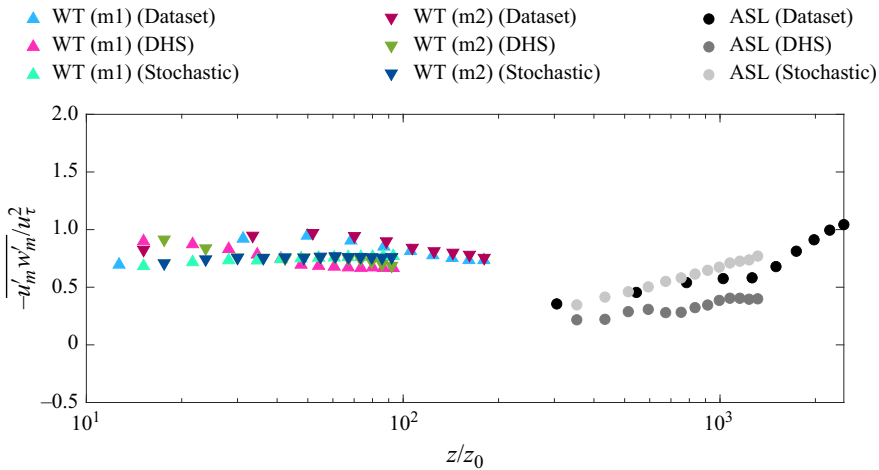

Figure 12. Reynolds shear stress profile of the PIV instantaneous velocity fields $-\overline{u'w'}$ (Dataset) compared with those generated using stochastic and DHS methods $-\overline{u'_m w'_m}$.

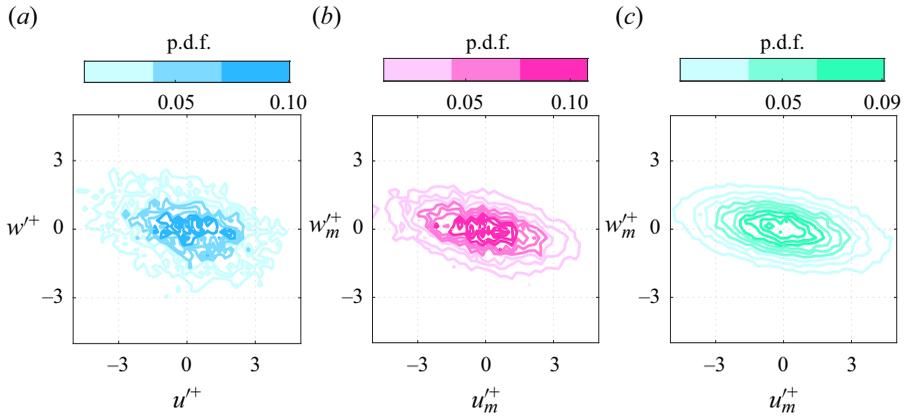

Figure 13. Quadrant-based analysis of Reynolds shear stress events at a given wall-normal position $z/\delta = 0.084$ for the WT (m1) dataset, based on (*a*) wind tunnel PIV instantaneous velocity ($u'-w'$) and (*b,c*) UMZ profiles ($u'_m-w'_m$) generated using the DHS method and the stochastic method, respectively.

Instantaneous Reynolds shear stress events are further mapped and categorised in the $u'-w'$ quadrant phase space (Wallace, Eckelmann & Brodkey 1972; Nakagawa & Nezu 1977; Raupach 1981; Wallace 2016). Figure 13 shows an example joint distribution of $u'$ and $w'$ at a given wall-normal position $z/\delta \approx 0.1$ for the wind tunnel WT (m1) dataset. Reynolds shear stress contributions are divided into low-momentum fluid travelling upward, referred to as 'ejection', and high-momentum fluid moving downward, known as 'sweep'. First, we recall that attached eddies, broadly overlapping with $\delta$-scale motions at laboratory-scale Reynolds numbers, are expected to provide a significant contribution to the Reynolds stresses, as discussed by Jiménez (1998), Christensen & Adrian (2001), Guala, Hommema & Adrian (2006) and Balakumar & Adrian (2007). As acknowledged previously, an independent uncorrelated stochastic generation of $u_m$ and $w_m$ UMZ attributes would not allow us to reproduce the expected Reynolds shear stress contributions and would thus display an isotropic quadrant event distribution. This limitation is partially overcome by the DHS method, which searches through the experimental dataset to find the most representative UMZ attributes given the initial generation of UMZ thickness $h_m$ and related mid-height position $z_m$ ($h_{m_i}$, $z_{m_i}$ based on the notation in figure 7). The signature of large-scale turbulence, expected to survive in the DHS-generated UMZs, provides a net contribution to the Reynolds shear stress for all datasets (figure 12), manifesting the typical $Q2$–$Q4$ alignment in the $u'-w'$ quadrants consistent with the experimental measurements (figure 13). The generation of two uncorrelated random numbers for $u_{m_i}$ and $w_{m_i}$ represent a case limit in a range of potential strategies to replicate the correlation between UMZ velocity component attributes, which are described in the discussion section. We opted for the generation of a set of $r_{u_{m_i}}$ $r_{w_{m_i}}$ numbers, imposing a uniform distribution $\in (0\ 1)$ and a prescribed $\overline{u'_m w'_m}$ correlation imposed by UMZs extracted from the experimental datasets. Results shown in figures 12 and 13 support our generation strategy and confirm the key role of UMZ spatial variability in the Reynolds shear stress.

Starting from the first UMZ thickness and modal velocity near the surface, the stochastic and DHS models keep generating the velocity layers of the step-like profile. Each velocity step $\Delta u_m$ represents the local difference between vertically adjacent modal velocities that are generated independently. The abrupt velocity jumps in the present model are a simplified representation of the internal shear layer (ISL), which have a finite thickness proportional to the Taylor microscale (Eisma *et al.* 2015; de Silva *et al.* 2017; Heisel *et al.* 2021). The ensemble average value $\Delta U_m$ is normalised with the shear velocity, as suggested by de Silva *et al.* (2017), and plotted in figure 14 for all datasets. The quantitative results are consistent with model predictions $\Delta U_m = 1.26u_\tau$ (Bautista *et al.* 2019) and previous experimental results (de Silva *et al.* 2017; Gul, Elsinga & Westerweel 2020; Heisel *et al.* 2020c). However, $\Delta U_m/u_\tau$ does not show the gentle decrease with the distance from the wall that has been experimentally observed and interpreted as a consequence of the diffusion of shear and vorticity and the progressive weakening of the ISLs (Heisel *et al.* 2021). Since there are no explicit modelling constraints on the velocity jump between generated UMZs, the observed collapse of $\Delta U_m/u_\tau$ is an emerging result of both models. It further confirms $u_\tau$ as a key velocity scale in the ISLs, well above the surface, throughout the whole outer region (Smits, McKeon & Marusic 2011). The underestimation of the DHS model near the wall, as compared with the stochastic model, is inherently related to the extracted modal velocities. As noted previously, small-scale flow features near the wall are not sufficiently represented or coherent to generate peaks in the velocity histograms, which are used to assign modal velocities. Hence, those near-wall features may be included in thicker UMZs characterised by higher velocities. The overestimate of modal velocities near wall, would also result in underestimation of the mean shear and, thus, of the velocity jump for the DHS model.

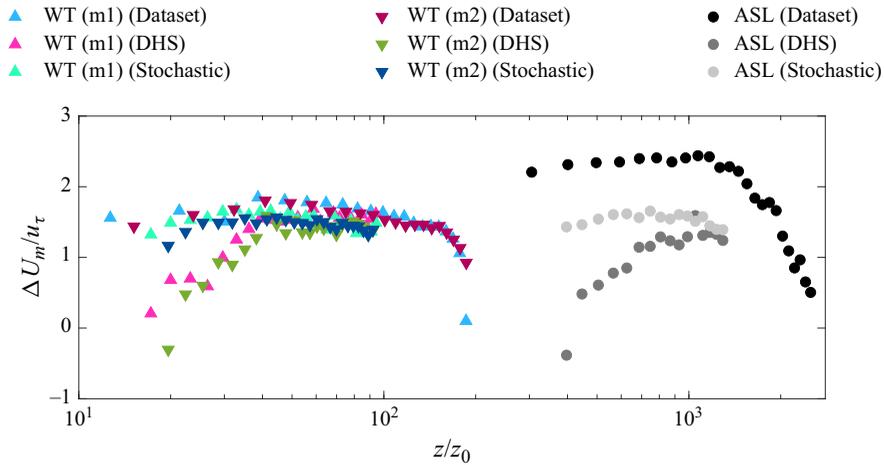

Figure 14. Profile of the average modal velocity jump $\Delta U_m$ across UMZ interface in the logarithmic region, normalised by the shear velocity. Comparison between experimental datasets and generated step velocity profiles.

Note that the modal velocity profile is expected to follow the logarithmic law of the wall. Hence, the log-normal distribution of UMZ thickness and the scaling of the velocity jump $\Delta U_m/u_\tau$ are deeply interconnected features. As a further element of validation, we show in figure 15 the joint p.d.f. of thickness $h_m$ and various elevations $z$, computed in the same manner as for figure 4. It can be seen that the stochastically generated UMZs thickness by both methods exhibit characteristics of wall-attached behaviour in the log region, comparable to the measured UMZ attributes (see figure 4). It is important to highlight that the generation of the UMZ thickness in both models is carried out stochastically using the same procedure. This holds true for both figures 15 and 16. The comparison between generated and measured data is further emphasised in figure 16 where the mean UMZ thickness is plotted as the function of the height. The emerging $h_m$, $z_m$ scaling relationship is expected to weaken outside of the log region, in the wake, where the constraint of the outer scale $\delta$ becomes increasingly relevant, as observed from other datasets included in Heisel *et al.* (2020*c*).

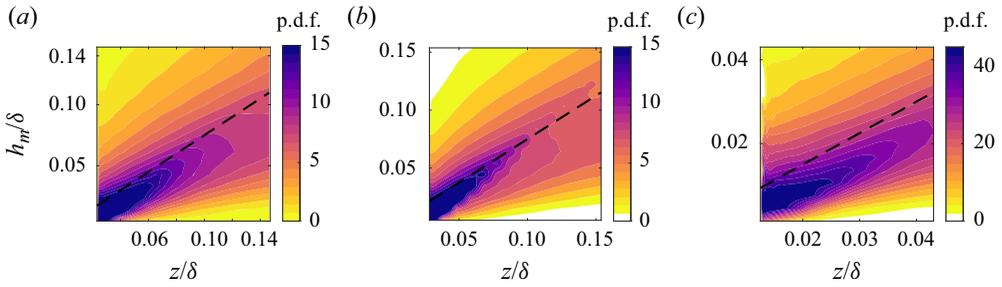

Figure 15. Joint p.d.f. of UMZ thickness $h_m$ and elevation $z$ from the generated velocity profiles for the wind tunnel $(a,b)$ and the ASL flows $(c)$; the dashed line indicate $H_m(z) = 0.75z$, highlighting wall-attached behaviour. The stochastic generation of UMZ thickness, in both models, is based on inverse transform sampling.

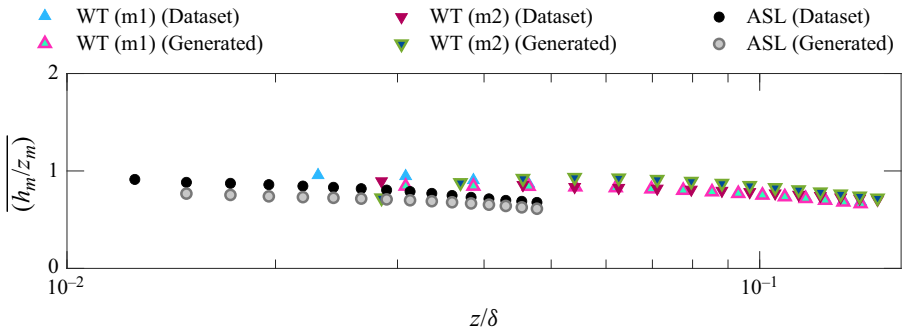

Figure 16. Profile of the average normalised UMZ thickness $\overline{(h_m/z_m)}$ for all datasets and corresponding stochastically generated profiles.

The coupling between UMZ modal velocity and thickness attributes can be further explored by estimating the local velocity gradient $\partial U/\partial z \simeq \overline{(\Delta u_m/h_m)}$. This emerges from the ensemble behaviour of the identified UMZs, specifically from the corresponding attributes averaged at each wall-normal location $z$: the velocity jump between nearby UMZs $\Delta u_{m_i}$ (figure 7*d*) and the mean of their thickness $h_m = (h_{m_i} + h_{m_{i+1}})/2$. In figure 17 the averaged modal velocity gradient, normalised by the friction velocity and the aerodynamic roughness length $\overline{(\Delta u_m/h_m)}(z_0/u_\tau)$, is obtained from the stochastic and DHS models and plotted for all datasets. As observed, the theoretical trend $(\partial U/\partial z)(z_0/u_\tau) = z_0/\kappa z$ is recovered at elevations above the roughness sublayer, where no criticalities are expected in identifying UMZs. Results from both models, in the absence of constraints on the modal velocity jump, confirm that the recovered logarithmic mean velocity profile (figure 10) is ingrained in the step-like structure of the UMZs. The velocity jump scaling with $u_\tau$ (figure 15) appear to be the two minimal ingredients for potentially predictive low-dimensional models.

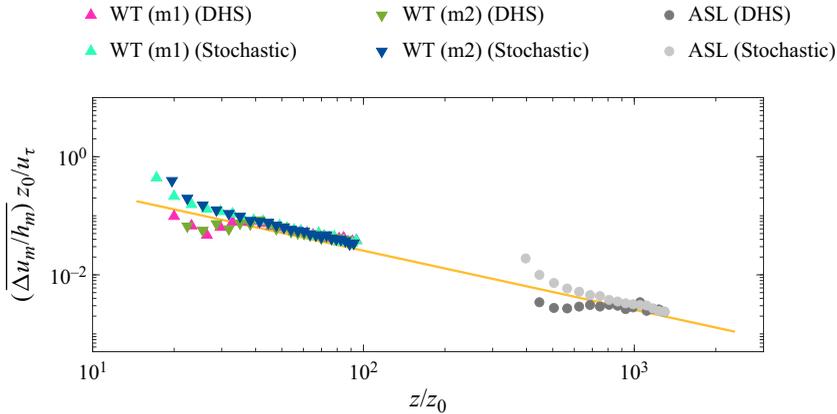

Figure 17. Average modal velocity jump gradient $\overline{(\Delta u_m/h_m)}$ profile, normalised with friction velocity $u_\tau$ and aerodynamic roughness length $z_0$. The yellow line marks the theoretical profile $z_0/\kappa z$, derived from the log law.

Note that the emerging close comparison with the mean velocity gradient, in the stochastic method, is preserved using both an independent and a coupled generation of random numbers for the c.d.f. inversion of $u_m$ and $h_m$. As expected it is also preserved through the nearest-neighbour procedure of the DHS method selecting UMZs with attributes, such as $u_m$, statistically consistent with the generated thickness $h_m$ at the prescribed height $z_m$.

## 5. Discussion

### 5.1. *Independent or joint generation of UMZ attributes?*

In this section we provide some experimental evidence on the mutual relationship between UMZ attributes, in particular on $u_m$–$h_m$ and $u_m$–$w_m$, both posing some questions on the optimal random number generation strategy. The stochastic method employed in the synthetic UMZ step generation has been tested following two simple scenarios: (i) a prescribed functional dependency between the random numbers $r_{u_{m_i}}$, $r_{h_{m_i}}$ used for the modal velocity and step height generation, respectively; and (ii) uncorrelated $r_{u_{m_i}}$, $r_{h_{m_i}}$ (used to produce current figures). Extracted attributes from identified UMZs, in the logarithmic regions of the three datasets, have been guiding our methodology and assumptions. For this specific question, we tuned the UMZ extraction algorithm to provide one single set of attributes per zone, assigned at a wall-normal coordinate $z_{ic}$ defined by the centroid of each UMZ. The extracted $h_m$, $u_m$ and $w_m$ are denoted as 'modal field'. The statistical description of UMZ's attributes is provided at every $z_i$, compiling all the zones extending in the vertical direction to include the reference height ($z_{ic} - h_m/2 < z_i < z_{ic} + h_m/2$). In figure 18 we illustrate the joint distribution of $u_m$–$h_m$ at $z_i/\delta \simeq 0.025$–0.12, for the WT (m1) wind tunnel identified UMZs (other datasets show qualitatively similar features). Close to the wall, the appreciably positive correlation is due to the occurrence of large high-momentum zones, i.e. with large $h_{m_i}$ and $u_{m_i}$ values in figure 18(*b*), that extend down near the wall. This observation is consistent with the prevalence of $Q4$ sweep events in the roughness sublayer (Heisel *et al.* 2020), and suggests potential $r_{u_{m_i}}$, $r_{h_{m_i}}$, and thus $u_m$–$h_m$, dependencies. However, with increasing $z_i$ the correlation is smeared, supporting the current adoption of uncorrelated $r_{u_{m_i}}$, $r_{h_{m_i}}$, reflecting the more even balance of $Q2$ and $Q4$ contributions in the logarithmic region. Surprisingly, the decreasing correlation trend survives in the synthetic generated profiles.

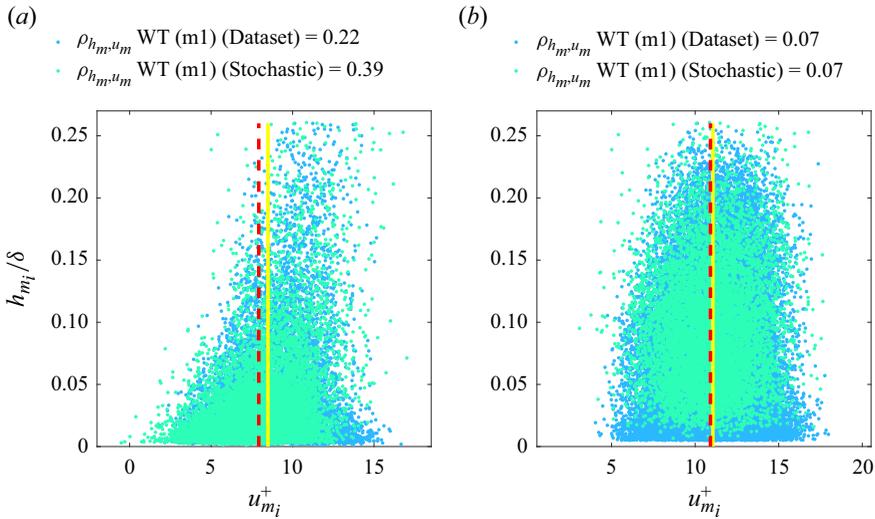

Figure 18. Joint distribution of the UMZ modal velocity and thickness extracted from the wind tunnel WT (m1) dataset, as single UMZ attribute (blue dots) and generated stochastically (green dots). Results are shown at different wall-normal positions (a) $z_i/\delta = 0.025$ and (b) $z_i/\delta = 0.122$. Yellow solid lines mark the mean modal velocity of the dataset; red dashed lines indicate the mean modal velocity value of the stochastically generated profiles. As the elevation increases, the correlation coefficient between the extracted $h_m$ and $u_m$ decreases.

The other potentially correlated UMZ attributes under exam are the modal $u_m$ and wall-normal averaged velocity $w_m$, which are, to some extent, expected to contribute to the Reynolds shear stress of the instantaneous field. We again rely on identified UMZs from the experimental datasets, using one set of attributes per zone, and compiled statistics at specific $z_i$. A level of correlation is expected (see figure 13), since uncorrelated $r_{u_{m_i}}, r_{w_{m_i}}$ lead to a non-physical $\overline{u'_m w'_m} = 0$. However, a perfectly anti-correlated $r_{w_{m_i}} = 1 - r_{u_{m_i}}$, while guaranteeing a uniform distribution and a correct inversion of the respective c.d.f., lead to an overestimate of the Reynolds shear stress contribution by UMZs. Figure 19 reports the $\overline{u'w'}$ profile for the instantaneous velocity filed, as well as for the experimentally collected modal field ($\overline{u'_m w'_m}$) for the different stochastic generation strategies. We show that imposing the UMZ specific correlation coefficient derived from the data allows us to reproduce the contribution of the modal field, which is only slightly lower than the actual Reynolds stresses. A perfect correlation leads to an overestimate of $\overline{u'_m w'_m}$ across the whole wall region. This result confirms that both extracted and simulated variability of the UMZ modal field accounts for a very significant portion of the Reynolds stress tensor $\overline{u'_i u'_j}$, including diagonal (variances) and off-diagonal (shear) terms. The latter comparison forced us to adopt a $r_{u_m}, r_{w_m}$ generator procedure ensuring (i) a uniform distribution for each of the two random numbers and (ii) a prescribed joint $r_{u_m} r_{w_m}$ distribution to recover the experimentally estimated correlation coefficient. These two random numbers $r_{u_m} r_{w_m}$ are generated from Gaussian copula with desired linear correlation coefficient $\rho_{r_{u_m}, r_{w_m}}$ in order to satisfy both mentioned required conditions (Joe 1997; Nelsen 2007).

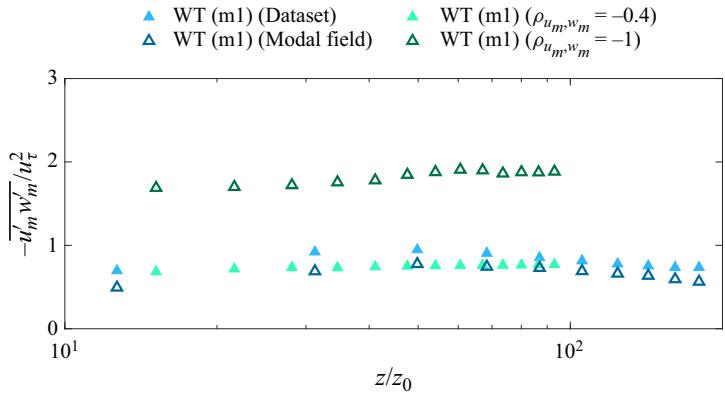

Figure 19. Vertical profile of the modal velocity contribution to the Reynolds shear stress. Comparison between experimental PIV velocity field $\overline{u'w'}$ (Dataset), extracted UMZ $\overline{u'_m w'_m}$ (Modal field) and stochastically generated profiles imposing different correlation coefficient values between $u_m$ and $w_m$.

### 5.2. Generalised modelling for a statistical description of UMZs

In this section we attempt a generalised, yet simplified, description of the parameters required for the stochastic generation of UMZ attributes at different wall distances within the logarithmic region. These generalisations leverage both the present experimental observations and known trends from the literature. Figure 5 showed the UMZ thickness and velocity distribution parameters fitted to the experimental data and required to accurately reproduce the p.d.f. and c.d.f. of the velocity step-like profiles. We remark that the thicknesses of the collected UMZs have a log-normal distribution (figure 3) at each given $z_i$ position, and should thus be normalised to be incorporated in the model in a logarithmic form. The wall-normal distance $z_i$ is a suitable length scale for normalising the thickness of the collected UMZs in the logarithmic region, as demonstrated in Heisel *et al.* (2020c) and confirmed in figure 4. The normalised profile of the mean and standard deviation of thickness, which are represented by $\hat{\mu}_{H_m}^{z_i} = \mu(\log(h_m/z_i))$ and $\hat{\sigma}_{H_m}^{z_i} = \sigma(\log(h_m/z_i))$, respectively, are plotted in figure 20($a$,$b$), with the abscissa being normalised by the outer length scale $\delta$. The discrepancy between the wind tunnel datasets and the ASL dataset may be attributed to uncertainty in determining the thickness of the boundary layer $\delta$ for the ASL dataset, or non-stationarity effects in the ASL also contributing to the $z_0$ and $u_\tau$ estimates. The profiles of the mean and standard deviation of the modal and vertical velocity of the UMZ, normalised with the friction velocity $u_\tau$, are plotted as a function of the aerodynamic roughness length $z_0$, and are shown in figures 20($c$)–20($f$). The standard deviation values for all UMZ attributes are approximately constant and independent of the wall-normal distance, whereas the mean velocity profile is assumed to obey to the logarithmic law. The mathematical formulation of the statistical moments of UMZ attributes required for the estimate of height-dependent p.d.f. and c.d.f., and ultimately for the generation of the synthetic modal velocity fields, can be simplified, within the range of Reynolds number explored here, as

$$\hat{\mu}_{H_m}^{z_i} \simeq -3.59 \left( \frac{z}{\delta} \right)^{0.91}, \quad \hat{\sigma}_{H_m}^{z_i} \simeq 1, \quad \mu_{U_m}^{+} \simeq \frac{1}{\kappa} \log \left( \frac{z}{z_0} \right),$$
$$\sigma_{U_m}^{+} \simeq 2, \, \mu_{W_m}^{+} \simeq 0, \quad \sigma_{W_m}^{+} \simeq 0.85. \qquad (5.1a\text{–}e)$$

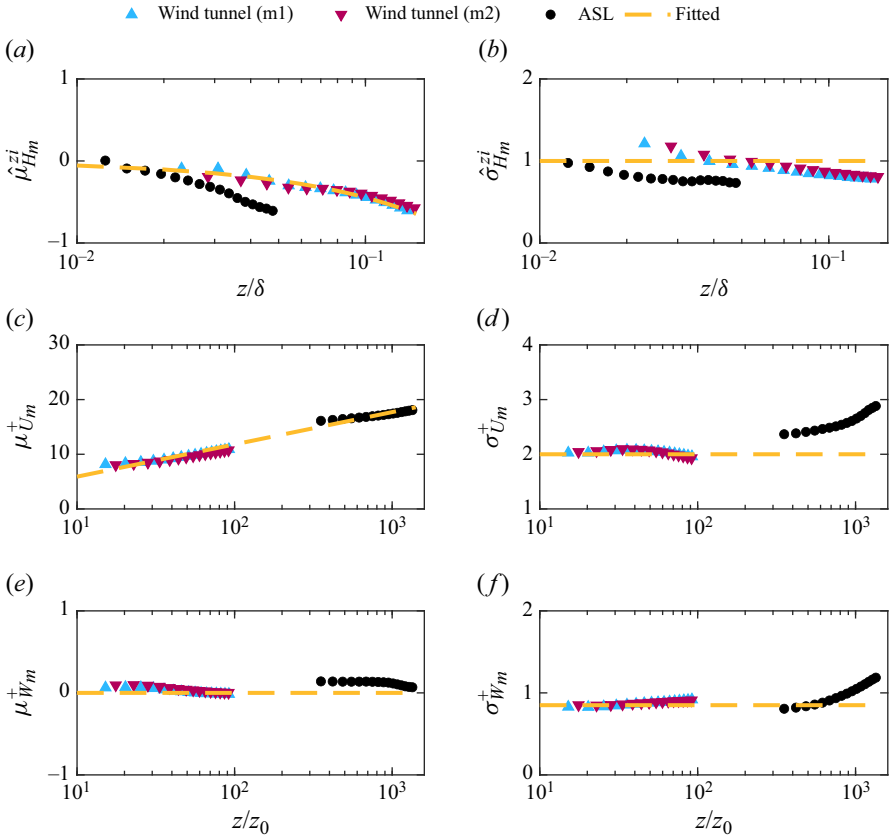

Figure 20. Required dimensionless parameters to reproduce p.d.f. and c.d.f. of UMZ attributes at different wall-normal positions, for the three datasets: (a) $\hat{\mu}_{H_m} = \mu(\log(h_m/z_i))$ is the mean of the logarithm of the thickness normalised by $z_i$; (b) $\hat{\sigma}_{H_m} = \sigma(\log(h_m/z_i))$ is the standard deviation of the logarithm of the thickness normalised by $z_i$; (c) $\mu_{U_m}^+$ is the mean modal velocity normalised by $u_\tau$; (d) $\sigma_{U_m}^+$ is the modal velocity standard deviation normalised by $u_\tau$; (e) $\mu_{U_m}^+$ is the mean of the wall-normal velocity normalised by friction velocity; (f) the standard deviation of wall-normal velocity $\sigma_{W_m}^+$ normalised by the friction velocity. Orange dashed lines are the simplified model prediction based on (5.1a–e).

The statistics extracted from the ensemble of generated velocity profiles do not exhibit significant differences from the figures shown previously. However, such a simplified description of the parameters profile is only valid in the logarithmic region of canonical, fully rough, turbulent boundary layers, and may be extended down to the roughness sublayer with some caution. Note that in the normalised description of mean and standard deviation of UMZ attributes, the equations for generating the thickness, modal and vertical velocity of UMZ are modified for the thickness $(h_{m_i} = z_i \exp(\hat{\mu}_{H_m}^{z_i} + \hat{\sigma}_{H_m}^{z_i}\sqrt{2}\mathrm{erfinv}(2r_{h_{m_i}} - 1)))$, and for the modal and vertical velocities $u_{m_i} = u_\tau(\mu_{U_m}^+ + \sigma_{U_m}^+\sqrt{2}\mathrm{erfinv}(2r_{u_{m_i}} - 1))$ and $w_{m_i} = u_\tau(\mu_{W_m}^+ + \sigma_{W_m}^+\sqrt{2}\mathrm{erfinv}(2r_{w_{m_i}} - 1))$.

## 6. Conclusion

In this study, we provide experimental results on UMZ attributes (thickness $h_m$, modal $u_m$ and vertical $w_m$ velocity) extracted from laboratory and field measurements of rough wall turbulent boundary layers, covering a broad range of Reynolds numbers ($Re_\tau \sim O(10^4 - 10^6)$) and surface roughness. Leveraging on these results, we propose two models able to generate step-like instantaneous velocity profiles that reproduce the statistical properties of the investigated canonical flows. The first model utilises a stochastic approach to determine all features of the UMZs contributing to the step-like velocity profile. The second model employs a stochastic generation of only the UMZ thickness, while utilising a data-driven technique to estimate the associated modal and vertical velocities (DHS). To perform stochastic-based calculations, the distributions of UMZ attributes must be known at each elevation $z$, so that the c.d.f. can be formulated and inverted. From the estimated distribution based on experimental data, we assumed that the p.d.f. of UMZ thickness is log-normal, whereas modal and vertical velocity are Gaussian. The required parameters of these distributions are the mean and standard deviation of UMZ attributes, which are extracted at each $z$ from the laboratory and field datasets. The statistical moments of the synthetically generated step-like velocity profiles are compared with those calculated from direct measurements in rough wall turbulent boundary layers. There are a few major outcomes from this work. The validated distribution functions of height ($z$)-specific UMZ attributes, presented in § 5.2, provide the opportunity to introduce variability in the UMZ step-like structure. The assumptions and emerging results allow the recovery from the synthetically generated streamwise velocity profiles of the logarithmic law of the wall and of the associated mean shear. These assumptions are consistent with the scaling of the AEH: the UMZ thickness $h_m$ scaling with $z$ and the velocity jump across the ISLs scaling with $u_\tau$. The variability of the modal velocity field, confined to the evolution of UMZs and ISLs, accounts for the full streamwise and vertical velocity variances, leaving only a limited portion of the Reynolds shear stress to the vortices or other under-represented high frequency fluctuations not featured in our models. Particular care is devoted to the coupling between UMZ attributes, such as the UMZ thickness, the streamwise modal and the wall-normal velocities. It appears that, for $h_m$ and $u_m$, using the data-driven model is equivalent to the fully stochastic UMZ generation, with no prescribed correlation between modal velocity and thickness, suggesting an approximate balance between $Q2$ and $Q4$ events in the logarithmic region. A non-zero correlation, imposed by the experimental modal velocity datasets, was however imposed in the $u_m$, $w_m$ generation to retrieve a consistent (modal) Reynolds shear stress contribution. Because of the entanglement between input parameters, dataset assimilation and model outcomes, we consider the emerging velocity jump scaling $\Delta U_m \sim u_\tau$ as the most convincing result for the model validation, also in view of the close agreement with previous studies (Bautista *et al.* 2019; Heisel *et al.* 2020*c*). The agreement between the mean shear from instantaneous UMZs $\overline{(\Delta u_m/h_m)}$ and the corresponding shear from the logarithmic law of the wall, is an important established link between the UMZ generation and the mean velocity profile. A second minor argument supporting the formulation of the model is the emerging log-normal distribution of $h_m(z)$, introducing the necessary skewness to enable the generation of relatively thick and fast zones near the wall, which seems to be a critical source of variability. The proposed models possess several advantages, such as the ability to introduce a structurally consistent variability in both velocity components, from a basic parameterisation of the averaged velocity profile, and the applicability to a wide range of Reynolds numbers and surface roughness. So far, the most serious limitations include the confinement of the current model to the logarithmic layer, due to a lack of observations in the wake region, and the two-dimensional description of the UMZ structure, due to a lack of simultaneous measurements in the vertical and wall-parallel planes.

**Funding.** This work was partially supported by the National Science Foundation (grant number 2018658).

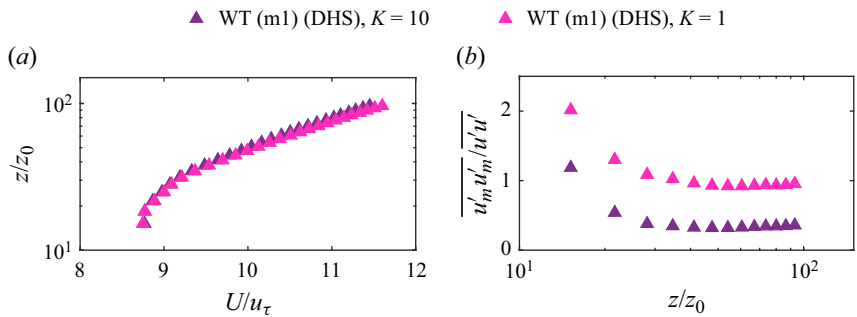

Figure 21. (*a*) Mean modal velocity profile and (*b*) streamwise modal velocity variance profile, estimated on DHS-generated velocity steps for the varying number of nearest neighbours $K$.

**Declaration of interests.** The authors report no conflict of interest.

**Data availability statement.** The wind tunnel data that support the findings of this study are openly available in the Data Repository for U of M at https://doi.org/10.13020/6grw-ny29.

**Author ORCIDs.**

 Roozbeh Ehsani https://orcid.org/0009-0005-9273-6342;

 Michael Heisel https://orcid.org/0000-0002-4200-5550;

 Jiaqi Li https://orcid.org/0000-0002-1201-7489;

 Jiarong Hong https://orcid.org/0000-0001-7860-2181;

 Michele Guala https://orcid.org/0000-0002-9788-8119.

## Appendix A

### A.1. *Parameter selection for the DHS model*

In order to discuss the hyper-parameter $K$ required in the DHS model, its impact on the reproduced first and second statistical moments of the modal velocity is studied. The hyper-parameter $K$ is the chosen number of neighbours, out of $O(10^6)$ samples, to determine the modal and vertical velocity of each generated UMZ, given $h_m$ and $z_m$. Figure 21(*a*) illustrates the sensitivity of the mean velocity profile, averaged over an ensemble of step-like modal velocities generated by the DHS model, for different values of $K$. The log-law can be approximately recovered with weighted averaging ($K = 10$) or without weighted averaging ($K = 1$) of the modal velocities of neighbouring UMZs, from our datasets. However, selecting $K = 10$, the weighted averaging results in the generation of synthetic modal velocity profiles that are more similar to each other and to the mean logarithmic velocity profile. This results in smaller deviations from the mean velocity at each $z_i$, and in the underestimation of the streamwise velocity variance profile, as shown in figure 21(*b*). The case limit is by represents $K = 1$, where for any given $h_m$ and $z_m$, the experimental dataset will yield the most accurate match for $u_m$. This implies that the variability observed in the measured and identified zones is entirely retained in the generated profiles, including the variance.